\documentclass[10pt,preprint2]{aastex}
\usepackage{wasysym}

\shorttitle{Wind-Driven Shapes of Transiting EGPs}
\shortauthors{Barnes, Cooper, Showman, \& Hubbard}

\begin{document}

\title{Detecting the Wind-Driven Shapes of Extrasolar Giant Planets from Transit 
Photometry}
\author{Jason W. Barnes}
\affil{Department of Physics}
\affil{University of Idaho}
\affil{Moscow, ID 83844-0903}
\email{jwbarnes@uidaho.edu}

\author{Curtis S. Cooper\altaffilmark{1}, Adam P. Showman, William B. Hubbard}
\affil{Department of Planetary Sciences}
\affil{University of Arizona}
\affil{Tucson, AZ  85721}
\altaffiltext{1}{also of NASA Astrobiology Institute}

\newpage

\begin{abstract}

Several processes can cause the shape of an extrasolar giant planet's shadow, as viewed in
transit, to depart from circular.  In addition to rotational effects, cloud formation,
non-homogenous haze production and movement, and dynamical effects (winds) could also be
important.  When such a planet transits its host star as seen from Earth, the asphericity
will introduce a deviation in the transit lightcurve relative to the transit of a perfectly
spherical (or perfectly oblate) planet.  We develop a theoretical framework to interpret
planet shapes.  We then generate predictions for transiting planet shapes based on a
published theoretical dynamical model of HD189733b.  Using these shape models we show that
planet shapes are unlikely to introduce detectable lightcurve deviations (those
$>~1\times10^{-5}$ of the host star), but that the shapes may lead to astrophysical sources
of systematic error when measuring planetary oblateness, transit time, and impact parameter.

\end{abstract}

 
 \section{INTRODUCTION}

Transits are proving to be the key to characterizing extrasolar giant planets. 
Transit lightcurve photometry has allowed measurements of planets' orbital
inclination and radius, which when combined with radial velocity observations allows
an unambiguous determination of a planet's mass and density.  To date such
measurements have been made for 55 transiting planets (for an up-to-date list of
known transiting planets see the Extrasolar Planets Encyclopedia at 
{\tt http://exoplanet.eu/catalog-transit.php/}).

Previous groups have suggested that visible-light transit photometry could determine a planet's 
oblateness \citep{Seager.oblateness,oblateness.2003}, the existence of ring systems
\citep{2004ApJ...616.1193B}, and/or the potentially artificial nature of the transiting object
\citep{2005ApJ...627..534A}.  However, the published lightcurves of each of the transiting
planets discovered thus far indicate no deviations from sphericity within errors.  Infrared
observations of transiting planets' secondary eclipse has revealed the dayside brightness
temperature of TRES-1 \citep{2005ApJ...626..523C}, HD209458b
\citep{2005Natur.434..740D,2008ApJ...673..526K} and HD189733b
\citep{2006ApJ...644..560D,2008ApJ...686.1341C,2008Natur.456..767G} and the spectrum of
HD209458b \citep{2007Natur.445..892R,2008ApJ...674..482S} and HD189733b
\citep{2007ApJ...658L.115G}.

Numerical simulations of atmospheric dynamics predict that the photospheres of hot Jupiters
are non-homogenous
\citep{2003ApJ...587L.117C,2005ApJ...629L..45C,2007ApJ...657L.113L,2008ApJ...673..513D,2008arXiv0809.2089S}. 
The longitudinal structure of these thermal variations was predicted to be detectable using
infrared photometry over the course of a full orbit \citep{2006ApJ...652..746F} and may also
be detectable during the planet's ingress and egress from secondary eclipse
\citep{2006ApJ...649.1020W,2007ApJ...664.1199R}.  Recent \emph{Spitzer} Space Telescope
measurements have shown phase-dependent infrared flux variability for $\mathrm{\upsilon}$
Andromedae \citep{2006Sci...314..623H} and HD189733
\citep{2007Natur.447..183K,2008arXiv0802.1705K}.  These measurements provide the first
constraints for dynamical models; however, the measured phase function does not match that
predicted theoretically.

In this paper, we show that atmospheric dynamics also introduces asphericity to a transiting
planet's observed shape, but that the lightcurve deviations thus produced are likely too small
to be detected.  First we investigate the processes that might affect a planet's shape (Section
\ref{section:shape}).  Next we derive an analytical expression for a planet's shape in the
simplified case of modification by a parameterized eastward equatorial jet (Section
\ref{section:analytical}).  We then develop a theoretical framework for the effects of planet
shape on transit lightcurves numerically using hypothetical regularly-shaped planets as a guide
(Section \ref{section:numerical}).  Finally we calculate the predicted shape of HD189733b based
on the new dynamical models of \citet{2008arXiv0809.2089S} (Section \ref{section:practical}),
and calculate the transit lightcurves for the predicted planet shape models, discussing shape
detectability from ground- and space-based photometry.

\section{PLANET SHAPE THEORY} \label{section:shape}

The radius of a planet viewed in transit corresponds to the radius at which the slant
optical depth of the planet's atmosphere is sufficient to block all potentially
transmitted light, which varies with the wavelength of observation
\citep{2000ApJ...537..916S}.  In general, this measured transit radius can differ
substantially from the nominal 1-bar-pressure radius used in planetary evolution
calculations due to molecular absorptions, clouds, Rayleigh scattering, or refraction
\citep{2001ApJ...560..413H,2003ApJ...594..545B,2002ApJ...572..540H}.  The long slant optical path through
a transiting planet's outer atmosphere leads to total optical depths between 35 and
90 times greater than the normal optical depth.  Hence seemingly unimportant
condensates or hazes can drive the transit radius to pressures as low as 1 millibar
\citep{2005MNRAS.364..649F}.  With an atmospheric pressure scale height near 500 km for
HD209458b \citep{2001ApJ...560..413H}, this effect would lead to a difference of
$\sim5000$ km, or $\sim5\%$ between the measured transit radius and the planet's
1-bar-pressure radius.

Most previous studies have treated transiting planets as either perfectly
spherical \citep[\emph{e.g.}][]{2001ApJ...560..413H} or perfectly oblate
\citep{Hui.Seager.2002,Seager.oblateness,oblateness.2003}.  However,
inhomogeneities in either the lateral or elevational distribution of molecular
absorbers or condensates (clouds and hazes) can lead to the departure of a
planet's shape from that of a sphere as measured in transit. 
\citet{2005A&A...436..719I} calculated that the day-night temperature contrast
on HD209458b could lead to an asymmetry in the abundance of sodium between the
morning and evening limbs of the planet as viewed in transit, though
\citet{2006ApJ...649.1048C} showed that carbon monoxide should be distributed
uniformly due to chemical disequilibrium effects.  

Three-dimensional dynamical simulations predict areas of upwelling, downwelling, and
horizontal jets on close-in extrasolar giant planets
\citep{2002A&A...385..166S,2005ApJ...629L..45C}.  Vigorous dynamics of the type
predicted are consistent with the formation of clouds if
condensibles are available and if the atmospheric pressure-temperature profile is
appropriate \citep{2000ApJ...538..885S,2003ApJ...586.1320C,2005ApJ...627L..69F}. 
However, to date no predictions of cloud cover patterns have been made for extrasolar
planets.

In addition to the possible creation of cloud bands, atmospheric dynamics can directly
affect a planet's transit shape by changing the constant-density surfaces.  At a given
gravitational equipotential surface, thermodynamic variations driven by insolation,
radiation, and air movement result in variations in air density.  High-speed jets are
particularly effective.  

Assuming that the atmospheric opacity is proportional to density, as would be the case if
the opacity resulted from refraction, Rayleigh scattering, molecular absorption, and
possibly condensate particulates, the $\tau=1$ height as a  function of azimuthal location
on the planet's disk would follow the constant-density surfaces.  If the constant-density
surfaces deviate from an equipotential surface, the deviant shape would alter the planet's
transit lightcurve.

\section{ANALYTICAL WIND-DRIVEN SHAPES}\label{section:analytical}

We quantify the dynamical shape deviations analytically to estimate the intensity of the
effects of a steady-state, idealized zonally symmetric jet.  We assume that the vertical
and north-south wind speeds are zero.  While realistic atmospheres will be decidedly more
complex, these assumptions allow us to derive an analytic toy model that will aid in
understanding the more realistic cases that we analyze in Section 5.

We start with the north-south component of the horizontal momentum equation from 
atmospheric dynamics
\citep[\emph{e.g.,}][]{Holton}: 
\begin{equation}
\frac{dv}{dt} = -fu - \frac{\partial \Phi(P, \phi)}{\partial y} - \frac{u^2 \tan\phi}{a}
\end{equation}
where $u$ is the eastward zonal wind speed, $v$ is the northward meridional wind speed,
$\phi$ represents the latitude, $a$ is the planet's radius, $y$ is the northward distance on
the sphere centered on the planet with radius $a$, $f$ is the coriolis parameter defined to
be  $f\equiv 2 \Omega \sin{\phi}$, $\Phi$ is the gravitational potential relative to that of
an oblate spheroid (\emph{i.e.}, not including the centrifugal term) and $\Omega$ is
the planet's rotation rate in radians per second.  Assuming a jet in gradient-wind balance
(\emph{i.e.} a 3-way balance between Coriolis, centrifugal, and pressure-gradient forces)
with an assumption of zero meridional wind (a good approximation even for solar system planets), then $\frac{dv}{dt}$ is zero, leaving us with
\begin{equation}
\frac{\partial \Phi}{\partial y} = -fu - \frac{u^2 \tan\phi}{a} .
\end{equation}
We then integrate to determine $\Phi(\phi, P)$, where $P$ refers to the atmospheric pressure,
\begin{equation}
\Phi = \int\frac{\partial\Phi}{\partial y}\mathrm{d}y
\end{equation}
where $y=a\phi$.  We assume a simple wind field $u(\phi)$ as a function of latitude 
($\phi$) with the winds zero
at the poles, $u_0$ at the equator, and varying sinusoidally in the mid-latitudes:
\begin{equation}
u=u_0\cos(\phi) .
\end{equation}
While not rigorously realistic, this north-south wind profile allows us to derive 
an analytic solution while
bearing at least superficial resemblance
to the superrotating equatorial jets predicted for hot Jupiters using 3D models 
\citep{2002A&A...385..166S,2005ApJ...629L..45C,2008arXiv0802.0327S,2008ApJ...673..513D}.  While
solar system planets have more complex zonal wind structures, hot Jupiters are predicted to have
a single broad jet similar to the one modeled.
Use of double-angle formulae, integration, and simplification results in:
\begin{equation}\label{eq:integration.result}
\Phi(\phi)-\Phi_0=\frac{u_0}{2}(\frac{u_0}{2}+\Omega a)
 \cos(2\phi')|_{\phi'=\phi_0}^{\phi'=\phi} .
\end{equation}
We assume that the jet is hydrostatically balanced and that the vertical temperature
profile is isothermal along the reference trajectory.  The isothermal assumption allows us to arrive at
an analytical solution, but we do not expect the shape of a non-isothermal planet to
differ significantly from the analytical solution thus derived.  We further assume that the gas behaves
ideally, \emph{i.e.} that $\rho=\frac{P}{RT_0}$.  
The constant $\Phi_0$ represents the reference potential at latitude $\phi_0$, 
\begin{equation}
\Phi_0=\Phi_\mathrm{ref}-RT_0\ln(\frac{P}{P_{\mathrm{ref}}}) .
\end{equation}
We take $\Phi_0$ to be equal to the true gravitational potential 
$\Phi_\mathrm{ref}$ when $\phi_0=0$ (\emph{i.e.}, the equator).
Evaluating Equation
\ref{eq:integration.result} then results in:
\begin{equation}\label{eq:integration.evaluated}
\Phi(\phi)=\frac{u_0}{2}(\frac{u_0}{2}+\Omega a)
 (\cos(2\phi)-1)+\Phi_0 .
\end{equation}
The effective gravitational potential as a function of
pressure ($P$) is then:
\begin{equation}
\Phi(\phi,P)=\frac{u_0}{2}(\frac{u_0}{2}+\Omega a)
 (\cos(2\phi)-1)+RT_0\ln(\frac{P}{P_\mathrm{ref}})-\Phi_\mathrm{ref} .
\end{equation}
Using the ideal gas law this becomes 
\begin{equation} 
\Phi(\phi, \rho) = \frac{u_0}{2}(\frac{u_0}{2}+\Omega a) 
(\cos(2\phi)-1)-RT_0\ln(\frac{\rho R T_0}{P_\mathrm{ref}})-\Phi_\mathrm{ref} .
\end{equation}

To solve for the magnitude of the 
resulting atmospheric-dynamics-induced shape change for the planet, we
look at the difference in magnitude of the polar and equatorial potentials,
\begin{equation}
\Delta\Phi \equiv \Phi(\phi=0) - \Phi(\phi=\frac{\pi}{2})~~\mathrm{.}
\end{equation}
Substituting in for the latitude values and assuming a constant-density surface we arrive at
\begin{equation}
\Delta\Phi = u_0 (\frac{u_0}{2}+\Omega a) \mathrm{,}
\end{equation}
which leads to an equator-to-pole radius difference of
\begin{equation}\label{eq:req-rp}
r_\mathrm{eq}-r_\mathrm{p} = \frac{u_0}{g} (\frac{u_0}{2}+\Omega a) 
\end{equation}
when we divide by the local acceleration due to gravity, $g$.  

For a jet on HD209458b of 1 km/sec, similar to the rotational speed at the planet's
equator, Equation \ref{eq:req-rp} predicts a modest wind-induced equator-to-pole
radius difference of 82 km.  A wind-induced shape change of this magnitude would be
comparable to that induced by the planet's rotation \citep{oblateness.2003}, and thus
difficult to discern for this (probably) tidally-locked planet.  

From numerical atmospheric circulation models, \citet{2005ApJ...629L..45C} simulate a robust
eastward jet for HD209458b with velocity of 2.8 km/s at the 20-bar pressure level, and 4
km/s at 220 millibars' pressure, similar to the results of other models
\citep{2008ApJ...682..559S}.  Substituting in these values for $u_0$ in Equation \ref{eq:req-rp}
yields wind-induced equator-to-pole radius differences of 480 km and 960 km respectively. 
These numerically-derived wind speeds would imply a planet shape difference comparable to
HD209458b's atmospheric scale height of 440 km \citep[\emph{e.g.},][]{2005MNRAS.364..649F},
and might therefore be detectable.   If the planet's slant optical depth
\citep{2005MNRAS.364..649F} is such that the pressure-level probed in transit is very high
in the atmosphere, the shape could be even more exaggerated.

\citet{2008arXiv0809.2089S} have more sophisticated coupled radiative-dynamical simulations
for extrasolar planet HD189733b.  These simulations show a qualitative similarity to those
of \citet{2005ApJ...629L..45C} in that they predict a strong west-to-east equatorial jet. 
The \citet{2008arXiv0809.2089S} model shows a maximum wind speed of $\sim3.5$ km/s at 10
millibars' pressure.  From this wind speed, equation \ref{eq:req-rp} predicts a
wind-induced equator-to-pole shape difference of 715 km for HD189733b at 10 millibars.

\section{PLANET SHAPE MEASUREMENT}\label{section:numerical}

The newly-launched \emph{Kepler} mission can measure lightcurves to strikingly high precision
\citep{2009Sci...325..709B}.  The measurement precision for the parent star of planet HAT-P-7b
(V=10.5) is $6\times10^{-5}$ per half-hour data point.  But the real power for detecting small
lightcurve deviations from \emph{Kepler} will come from the coaddition of a whole sequence of
transits.  In the HAT-P-7b case, for instance, there will be a total of 580 transits during the
\emph{Kepler} prime mission because of the planet's 2.2-day orbital period.  By coadding those
into a single lightcurve, the effective precision can be reduced by a factor of 24 to just
$2.5\times10^{-6}$ per half-hour measurement, or alternatively about $1.4\times10^{-5}$ for each
1-minute measurement (assuming that this star is put onto the short-cadence list).  Hence
lightcurve deviations at the 10 parts-per-million level may eventually be detected by
\emph{Kepler} by its end of mission.

In order to determine whether or not wind-induced departures from sphericity should be
detectable, we now numerically approach the problem of the effects of shape on
transit lightcurves.

\begin{figure} 
\includegraphics[totalheight=0.3\textwidth]{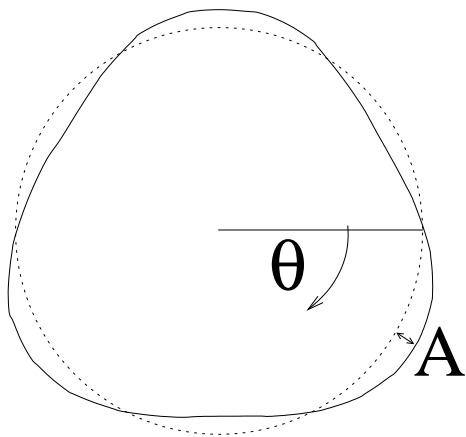}
 \caption{Geometry of shape parameterization. \label{figure:thetashape}}
\end{figure}

The lightcurves of all of the transiting planets known to date are well-modeled by
fitting just four values (assuming a value for the stellar mass):  the stellar radius
($R_*$), the planetary radius ($R_p$), the transit impact parameter ($b$, the distance
between the projected centers of the planet and star at mid-transit, in units of $R_p$),
and one or more parameters describing stellar limb darkening (\emph{e.g.}, $c_1$,
\citep{oblateness.2003}).  If a planet is not a perfect sphere, and if the deviation is
large enough, then its transit lightcurve would not be adequately modeled with this type
of 4-parameter fit.  Hence the residual from the 4-parameter spherical planet fit provides
a measure of the photometric detectability of deviations from a spherical planet
\citep{oblateness.2003}.

Deviations from either planetary sphericity \citep{Seager.oblateness,
oblateness.2003,2004ApJ...616.1193B,2005ApJ...627..534A} or uniform orbital motion
\citep{2007PASP..119..986B} lead to lightcurve residuals that are largest during planets'
transit ingress and egress.  Dynamically-driven deviations ought to be most
detectable near ingress and egress as well.

To establish a systematic framework within which realistic deviations can be
understood, we first investigate the detectability of hypothetical planets with
regularly-varied shapes.  Specifically we look at planets whose azimuthal profiles as
seen in transit vary as
\begin{equation}
r(\theta)=R_p+A\sin(n\Theta+\Psi)\label{eq:nshape}
\end{equation}
where $r$ is the projected planetary radius as a function of the azimuthal angle
$\Theta$ and $A$ is a constant with dimensions of length that corresponds to the 
amplitude of the deviation (Figure \ref{figure:thetashape}).  The parameter $\Psi$
represents a phase factor and affects the orientation of the resulting figure.  The 
positive-integer-valued $n$ allows for the generation of planets with varying numbers
of crenulations.  

The set of all shapes $r_n(\theta)$ with $\Psi=0, \pi/2$ provides an orthonormal basis that spans the
space of possible planet shapes.  Since any arbitrary planet shape can be represented
as a sum of shapes of the form $r_n(\theta)$, understanding the lightcurves of
these basic shapes will elucidate the nature of the lightcurves of more complex
objects.

We calculate transit lightcurves using the method of \citep{2004ApJ...616.1193B}:  an
explicit numerical integration of the light blocked by the planet relative to the
total stellar flux.  This calculation is makes no approximations regarding stellar
limb darkening.  Only computational time and double-floating-point precision limit its
accuracy.  We fit these artificial lightcurves with a spherical planet model using a
Levenberg-Marquardt algorithm to arrive at a least-squares-minimizing fit.  As in
\citet{2004ApJ...616.1193B}, we fit for the parameters $R_*$, $R_p$, $b$, and the limb
darkening parameter $c_1$ as defined in \citep{2001ApJ...552..699B}.

\begin{figure} 
\includegraphics[totalheight=0.55\textheight]{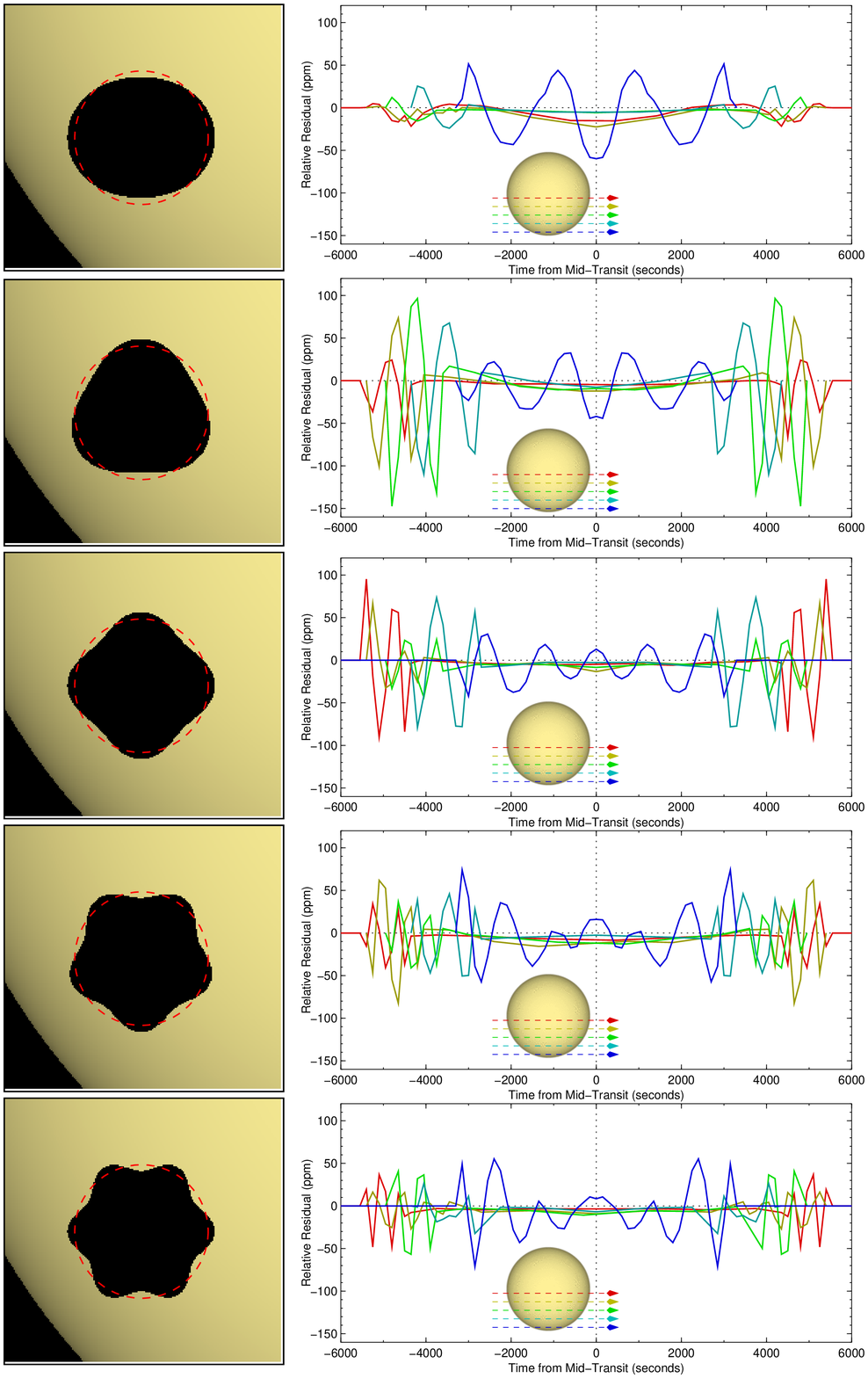}
 \caption{Lightcurve fit residuals for regularly-shaped planets that are symmetric
 with respect to their orbit normal.  From top to bottom they have (see Equation
 \ref{eq:nshape}):  $n=2$, $\Psi=\pi/2$;  $n=3$, $\Psi=0.$;  $n=4$, $\Psi=\pi/2$;  
 $n=5$, $\Psi=0$; and $n=6$, $\Psi=\pi/2$.\label{figure:symm}  The color for each
 lightcurve residual is colored such that red corresponds to impact parameter $b=0.1$,
 yellow to $b=0.3$, green to $b=0.5$, cyan to $b=0.7$, and blue to $b=0.9$.  The flux
 show is plotted relative to the total stellar flux.  This hypothetical planet has a
 radius equal to that of Jupiter and orbits a $1~M_\odot$ star with the same limb
 darkening as HD209458b with a semimajor axis of 1 AU.  To convert to other semimajor
 axes, multiply the x-axis by $sqrt{a_p}$ where $a_p$ is the planet's semimajor axis
 in astronomical units.  The value of $A$ used here is
 $0.1~R_\mathrm{Jup}$.  The amplitude of the transit residual is proportional to $A$
 for $A<<R_\mathrm{p}$.}
\end{figure}

\begin{figure} 
\includegraphics[totalheight=0.55\textheight]{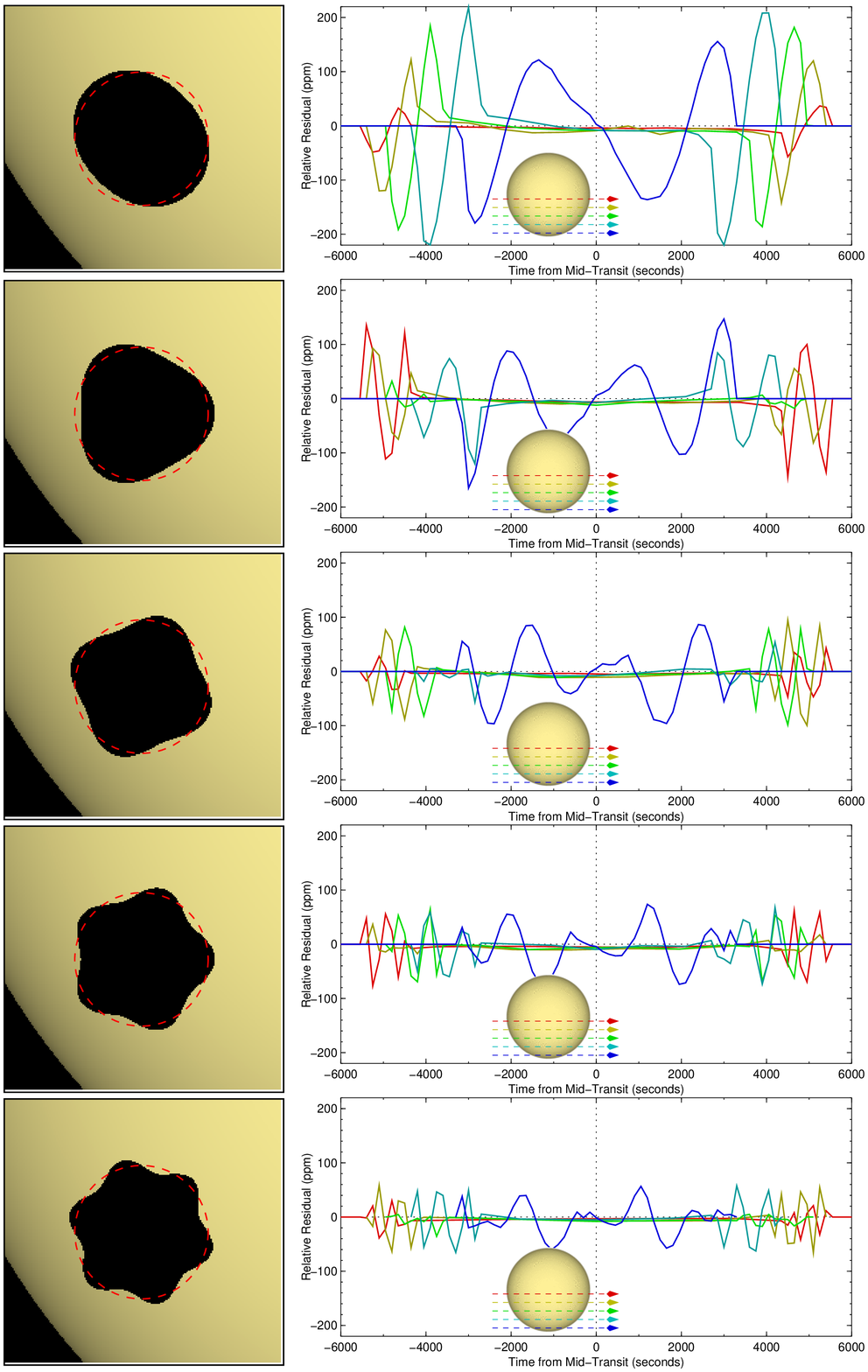}
 \caption{Lightcurve fit residuals for regularly-shaped planets that are asymmetric
 with respect to their orbit normal.  From top to bottom they have (see Equation
 \ref{eq:nshape}):  $n=2$, $\Psi=0$;  $n=3$, $\Psi=\pi/2$;  $n=4$, $\Psi=0$;  
 $n=5$, $\Psi=\pi/2$; and $n=6$, $\Psi=0.$  The color for each
 lightcurve residual is colored such that red corresponds to impact parameter $b=0.1$,
 yellow to $b=0.3$, green to $b=0.5$, cyan to $b=0.7$, and blue to $b=0.9$.  
 To convert to other semimajor
 axes, multiply the x-axis by $sqrt{a_p}$ where $a_p$ is the planet's semimajor axis
 in astronomical units.The value of $A$ used here is
 $0.1~R_\mathrm{Jup}$.  The amplitude of the transit residual is proportional to $A$
 for $A<<R_\mathrm{p}$.\label{figure:asymm}}
\end{figure}

The lightcurve residuals, after having been fit with the spherical-planet model, for
hypothetical transiting planets with $n=2,3,4,5,6$ are shown in Figures
\ref{figure:symm} and \ref{figure:asymm}.  Figure \ref{figure:symm} shows the
residuals for planets that are symmetric around a vector corresponding to their orbit
normal; Figure \ref{figure:symm} shows planets that are asymmetric with respect to
their orbit normal.  Symmetric planets that transit generate lightcurves that are
symmetric with respect to the mid-transit time.  The asymmetric planets generate
antisymmetric lightcurves due to the nature of their asymmetry.

The $n=1$ case (not shown) corresponds to planets whose center-of-mass is offset from
their projected center-of-figure.  The best-fit spherical planet model parameter for
this case have small systematic errors resulting from the offset, but their lightcurve
residuals are negligible.

The two-lobed case ($n=2$) forms an opaque ellipse in projection.  This kind of shape
might resemble that formed by a giant planet with a single strong prograde equatorial
jet.  In its transit lightcurve, such a planet behaves like an oblate planet. 
\citet{oblateness.2003} showed that these planets show a distinct transit residual
with one positive and one negative peak during ingress and egress.  Oblate planets
show a reduced signature for symmetric transits (Figure \ref{figure:symm}) relative to
oblate planets with antisymmetric transits due to variations in the spherical model fit
parameters that act to mimic the oblate transit signature.

Residuals for shapes with $n>2$ become progressively more complex.  These higher-order
shapes show a number of positive and negative deviations during ingress and egress. 
Empirically, as can be seen in Figures \ref{figure:symm} and \ref{figure:asymm}, the
number of peaks in the residual during planet ingress and egress is equal to $n$.  As
the length of ingress and egress is not a function of $n$, the duration of each
deviation becomes progressively shorter as $n$ increments.  Hence complex,
higher-order shapes would require a faster cadence for photometric observations in
order to be detected.

In the antisymmetric case (Figure \ref{figure:asymm}), the amplitude of the peaks in the
lightcurve residual decreases with increasing $n$.  Hence detecting higher-order
shapes requires better photometric precision than lower-order ones.  The symmetric
case is similar, with the exception of $n=2$ which has relatively low detectability due to the
spherical planet fit, as described above.

\begin{figure} 
\includegraphics[totalheight=0.135\textheight]{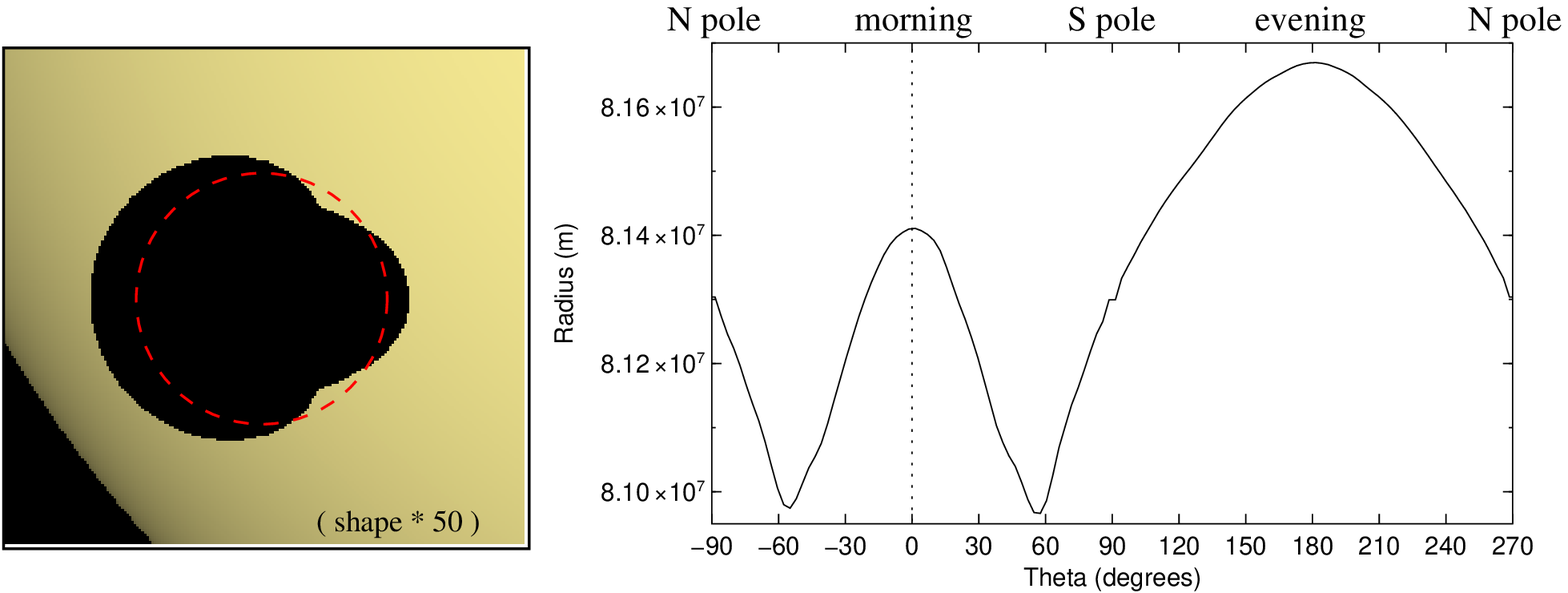}
 \caption{Predicted wind-driven shape of HD189733b presented as an image (left;
 exaggerated 50 times) and as a graph (right).  The average radius has been
 forced to the measured value, $1.138~\mathrm{R_{Jup}}$.  The dashed red line at left
 corresponds to the location and size of a spherical $1.00~\mathrm{R_{Jup}}$ planet 
 with the same center of mass as HD189733b for purposes of comparison.
 \label{figure:realplanetshapes}}
\end{figure}

\section{APPLICATION TO HD189733b} \label{section:practical}

Non-spherical planet shapes could be detectable with transit photometry, then, given a large
enough amplitude for the deviations from sphericity.  To test whether the effect should be
seen for known extrasolar planets, we estimate the shape of transiting planets HD189733b
using the \citet{2008arXiv0809.2089S} SPARC/MITgcm atmospheric model.   To arrive at a
planet shape, we assume that the slant optical depth of the planet's atmosphere is unity
where the density is $2.2\times10^5 \mathrm{~kg~m^{-3}}$.  The resulting shape is shown in
Figure \ref{figure:realplanetshapes}.

The SPARC model predicts a total variation in shape of $\sim700$~km.  This value agrees
very well with the theoretical value derived in Section \ref{section:analytical} (715 km).

In order to predict the lightcurve effects of this model-predicted shape, we create a
simulated transit lightcurve that we then fit with a spherical-planet model, as we did for
hypothetical regularly-shaped planets in Section \ref{section:numerical}.  The lightcurve
residuals that result are decidedly too small to be measured -- less than 1 part in $10^5$
of the stellar flux.  

To understand why the wind-driven shape is so hard to detect, we compare the model
HD189733b shape with the numerical results from Section \ref{section:numerical}.  Since for
small deviations from a sphere the detectabilities of various planet shapes are linear with
respect to the shapes themselves, the detectability for the model HD189733b should be the
same as a linear combination of the regular planet shapes from Section
\ref{section:numerical}.  We determine the relative contributions of each regular shape by
taking the Fourier transform of the predicted model shape shown in Figure
\ref{figure:realplanetshapes}.  The results are in Table \ref{table:fourier}.

\begin{deluxetable}{l|rr}
\tablecaption{Fourier components of model HD189733b shape from Figure 
\ref{figure:realplanetshapes}.\label{table:fourier}}
\tablewidth{0pt}
\tablehead{
\colhead{Fourier} &
\colhead{Symmetric} &
\colhead{Antisymmetric} \\
\colhead{Component} &
\colhead{Amplitude} &
\colhead{Amplitude}
}
\startdata
$n=0$  & $81360$ km &  -- \\
$n=1$  &$120.12$ km & $ 0.74$ km\\
$n=2$  & $-0.51$ km &$-64.78$ km \\
$n=3$  & $55.68$ km & $-0.31$ km\\
$n=4$  & $-0.51$ km & $32.45$ km \\
$n=5$  & $-4.61$ km & $-0.25$ km \\
$n=6$  & $-0.30$ km &  $7.66$ km \\
\enddata
\tablecomments{There is no antisymmetric component for $n=0$, which corresponds to the
average spherical radius.  We define the $n=1$ shape component with $\Psi=0$ 
as ``symmetric" to be consistent with the higher order components (see Figure 2, caption).
What is listed as the ``antisymmetric" component of 
$n=1$, the $\Psi=\pi/2$ component,
actually produces a symmetric lightcurve as well, but is shown under that heading
for consistency with the higher order shapes.  See text.  That all of the $\Psi=\pi/2$
components have so little power owes to the planet's symmetry around its own equator.} 
\end{deluxetable}

While the dynamical model predicts the total equator-to-pole radius difference to be of
order $700$~km (which would correspond to an amplitude of 350 km from Table
\ref{table:fourier} if it were all in one component), the Fourier components in Table
\ref{table:fourier} show that the majority of the amplitude is in the $\Psi=\pi/2$
symmetric components.

The $n=1$ component corresponds to an offset between the planet's center of mass and its
center of figure.  This component is inherently undetectable using transit photometry
alone, as it is sensitive only to the figure and not to the planet's mass.  The $n=2$
symmetric component corresponds to the planet's oblateness.  As shown in Figure
\ref{figure:symm} (and in \citet{oblateness.2003}), this component has a relatively low detectability
owing to the ability of a spherical planet model to partially emulate the oblate
planet's transit lightcurve.  The higher-order components, $n=3, 4, 5, 6$, have
progressively lower inherent detectabilities and lower Fourier amplitudes in the dynamical
model prediction as well, leading to their having very small effects on the planet's
transit lightcurve.

Though the raw wind speeds predicted for hot Jupiters would appear to lead to detectable
shape amplitudes, the nature of the shapes that result are such that the detectability is
low.  Hence wind-driven shapes of extrasolar planets are unlikely to be detectable in
transit photometry barring unusual high-velocity zonal wind structures.  It remains
possible that clouds or spatial variations in the intensity of haze or molecular absorption
could create unusual shapes with amplitudes large enough to be detected in some cases. 

If a transiting planet had a nonzero obliquity, then even a symmetrically uniform wind
jet would introduce a much more detectable antisymmetric lightcurve.  However, in order for a
planet not to have its obliquity reduced to near zero by tides, that planet would
necessarily need to be far from its parent star.  At those large distances, the stellar
insolation available to drive winds is lower, and hence we might expect a smaller amplitude
for the wind-driven shape (though nobody has modeled such a situation yet).  The net
detectability may be higher, though, due to the presence of antisymmetric Fourier
components.

\section{CONCLUSION}\label{section:conclusion}

Planetary winds affect the three-dimensional shape of a planet's  constant-density
contours, leading to departures from sphericity.  A planet's silhouette as viewed in
transit should depend on the projected shape of these constant-density surfaces, assuming
that absorption depends only on atmospheric density.  The resulting silhouette should
affect the planet's transit lightcurve.

We calculate an analytical estimate for the amplitude of the shape variation, using
simplifying assumptions regarding a planet's zonal wind structure.  Using the maximum
equatorial winds found in dynamical models of HD209458b \citep{2005ApJ...629L..45C} predicts
a atmospheric-dynamics-driven global equator-to-pole radius difference of between 480 km and
960~km.  Using results from the more sophisticated \citet{2008arXiv0809.2089S} model of
HD189733b, the analytic expression predicts a 715~km atmospheric-dynamics-driven global
equator-to-pole radius difference.

For a more robust estimate of a planet's transit shape, we use the
\citet{2008arXiv0809.2089S} model directly by deriving the wind-induced shape from contours
of constant density along the planet's terminator.  The resulting shape has a
total radius difference of around $\sim700$~km.  However, when we use our
numerical lightcurve-fitting routine to estimate the detectability of this shape in a
transit lightcurve, we find that it is not detectable, with lightcurve residuals of order
only $10^{-5}$ of the stellar flux.

To understand why the detectability is so low, we calculate detectabilities for planets
with regular shapes such that $r(\theta)=R_p+A\sin(n\Theta+\Psi)$.  The $n=0$ term
corresponds to the planet's average radius, the $n=1$ term to offsets between the center
of mass and center of figure, the $n=2$ term to planetary oblateness, and higher order
terms to more complex shapes.  The $n=1$ term is not detectable in a transit lightcurve. 
The projected shapes that are symmetric with respect to the planet's orbit normal produce
symmetric lightcurves; those shapes that are asymmetric produce antisymmetric lightcurves. 
In general shapes with lower $n$ have higher detectabilities, and require less-fine time
resolution than shapes with higher $n$.  However the symmetric $n=2$ term has somewhat 
low detectability due to the ability of a spherical planet model to partially mimic its 
transit lightcurve signature.

A Fourier decomposition of the model HD189733b shape reveals why it would be so hard to
detect.  Most of the amplitude of the shape is in the $n=1$ and $n=2$ terms:  the $n=1$ is
undetectable, and the $n=2$ has relatively low detectability.  The higher order terms become
progressively harder to detect as $n$ increases.  Thus the smooth nature of the
predicted HD189733b shape leads to its low transit detectability.

The detectability of other planets will depend on their wind velocities and zonal wind
structures.  However, the high insolation and low rotation rates of all hot Jupiters may
drive them to behave similarly to HD189733b.  Multiple counterrotating jets would avoid the
low detectability of the oblateness ($n=2$) term, but higher $n$ terms have lower
detectabilities as well.  Variation in the height of clouds or the absorption from haze or
the atmosphere around the disk could lead to detectable shapes, but only in unusual
circumstances.  Hence we think that it is unlikely that the winds on transiting planets
will be able to affect their lightcurves at a level that will be detectable in the
near-future. 

\acknowledgements

JWB and CSC were supported for this work in part by the NASA Postdoctoral Program,
administered by Oak Ridge Associated Universities through a contract with NASA, at Ames
Research Center and the LAPLACE Astrobiology Institute at the University of Arizona,
respectively.  APS was supported by a grant from the NASA Origins program.  JWB
acknowledges the Department of Physics at the University of Idaho for providing
publication costs for this paper via startup funds.

\bibliographystyle{apj}
\bibliography{references}

\newpage

\end{document}